# Isoelectronic Ru substitution at Fe-site in Sm(Fe$_{1-x}$Ru$_x$)As(O$_{0.85}$F$_{0.15}$) compound and its effects on structural, superconducting and normal state properties

M. Tropeano[1], M.R.Cimberle[2], C. Ferdeghini[1], G. Lamura[1], A. Martinelli[1], A. Palenzona[1,3], I. Pallecchi[1], A. Sala[4], and M. Putti[1,4]

[1]*CNR-SPIN, Corso Perrone 24, 16152 Genova, Italy*
[2]*CNR-IMEM, Dipartimento di Fisica, Via Dodecaneso 33, 16146 Genova, Italy*
[3]*Dipartimento di Chimica e Chimica Industriale, Università di Genova, Via Dodecaneso 31, 16146 Genova, Italy*
[4]*Dipartimento di Fisica, Università di Genova,Via Dodecaneso 33, 16146 Genova, Italy*

F.Bernardini[5], M. Monni[6] and S.Massidda[5]

[5]*Dipartimento di Fisica, Università di Cagliari, 09042 Monserrato (CA), Italy*
[6]*Istituto Officina dei Materiali del CNR (CNR-IOM) Unità SLACS, Cittadella Universitaria, 09042 Monserrato (CA), Italy*

I. Sheikin

*GHMFL, CNRS, 28 Avenue des Martyrs, BP 166, 38042 Grenoble, France*

## Abstract

In this work we present a *systematic* experimental and theoretical study of the structural, transport and superconducting properties of Sm(Fe$_{1-x}$Ru$_x$)As(O$_{0.85}$F$_{0.15}$) polycrystalline samples as a function of Ru content (*x*) ranging from 0 to 1. The choice of Ru as isoelectronic substitution at Fe site of F-doped compounds allows to better clarify the role of structural disorder in modifying the normal and superconducting properties of these newly discovered multiband superconductors. Two different regions are identified: the Fe-rich phase ($x<0.5$) where superconducting and normal state properties are strongly affected by disorder induced by Ru substitution; the Ru-rich phase ($x>0.5$) where the system is metallic and strongly compensated and the presence of Ru frustrates the magnetic moment on Fe ions. Here the lack of magnetic features and related spin fluctuations may be the cause for the suppression of superconductivity.

## 1.Introduction

The recent discovery of high critical temperature superconductivity in iron based compounds [1] has attracted a great deal of attention as these compounds appear to be a glaring case of proximity between superconductivity and magnetisms. The parent compounds exhibit antiferromagnetic spin-density-wave (SDW) order that disappears upon doping, giving rise to superconductivity. It has been suggested by many authors that superconductivity in pnictides could be mediated by magnetic excitations which couple electron and hole pockets of the Fermi surface, favoring s-wave order parameters with opposite sign on different sheets of the Fermi surface (s$^\pm$ coupling). [2]
The interplay between superconductivity and magnetisms can be investigated by varying magnetic and superconducting properties of the compounds through suitable substitutions. Moreover, scattering induced by substitutions is expected to affect superconductivity in very differently ways in the cases of conventional or unconventional coupling.[3] As a consequence, a thorough study of the behavior of T$_c$ vs structural disorder is crucial in order to probe different theoretical models.
Similarly to cuprates, the pnictide compounds have a layered structure characterized by the stacking of insulating and FeAs-conducting layers with general formulas *RE*FeAsO (*RE* being a rare earth) and $M$Fe$_2$As$_2$ ($M$ being alkaline/alkaline-earth metal) that have been indicated as 1111 and 122 families respectively.

Superconductivity emerges upon doping of either electrons or holes that can be realized in several ways, depending also on the system structure. In the 1111 family compounds, doping is realized by chemical substitutions in the charge reservoir layer by means of fluorine-oxygen substitution,[1] oxygen-vacancies [4], heterovalent substitution at the $RE$ site [5]. However, chemical substitution in the FeAs layer has also been proposed. [6,7] In particular substitution at Fe site is able to address two important questions at the same time: the evolution of superconductivity and magnetism with doping (and their interplay) and the robustness of the magnetic and/or superconducting order against substitutional disorder. In order to get reliable information on these central topics, it becomes crucial to separate them.

Several studies have been reported on 1111 compounds in which Fe is substituted by Co. The SDW order in the parent compounds is rapidly suppressed by Co doping, and superconductivity emerges at around $x$=0.025-0.05, shows a domelike behavior and disappears around $x$=0.2. [6,7] The normal-state resistivity exhibits semiconducting like behavior, making the Co-doped superconductors different from the F-doped ones, as a consequence of the role of disorder induced by the substitution in the conducting layer.

Also in F-doped superconducting 1111 compounds [8,9,10] Co doping increases the resistivity and about 10% of Co suppresses superconductivity, which can be an effect of overdoping and/or scattering due to substitutional disorder in the conducting layer.

Isoelectronic Ru substitutions of Fe have been investigated as well. In the 122 family the Ru substitution in the parent compounds, suppresses the SDW and superconductivity emerges in a very similar way as in Co doped systems.[11,12,13] These results indicate that also isoelectronic substitutions do affect the electronic structure, and significantly contribute to changes in carrier concentration and density of states at $E_F$. A different situation has been found in Pr-1111 where Ru substitution suppresses progressively SDW without the induction of superconductivity [14]. Finally Ru substitution in superconducting Nd-1111 has shown a lower tendency of decreasing $T_c$ than in Co doped compounds.[15]

All these results point out on the difficulty to separate the effects of disorder, doping and magnetisms on the superconducting properties of pnictides. The presence of multiple bands crossing the Fermi level makes the interpretation of data even more difficult and no conclusive studies have been reported to clarify the role of substitutional disorder in superconducting pnictides. For this reason, we present a *systematic* experimental and theoretical study of the structural and transport properties of Sm(Fe$_{1-x}$Ru$_x$)As(O$_{0.85}$F$_{0.15}$) samples as a function of Ru content ($x$) ranging from 0 to 1. Theoretical calculation allows to better clarify the role of Ru as isoelectronic substitution at Fe site. A deep investigation of transport properties in normal and superconducting state allows to separate the roles of disorder and magnetisms on the superconducting properties of these multiband superconductors.

## 2. Synthesis and structural characterization

Polycristalline samples of Sm(Fe$_{1-x}$Ru$_x$)As(O$_{0.85}$F$_{0.15}$) were synthesized starting with SmAs (pre-synthetized), Fe$_2$O$_3$, RuO$_2$, FeF$_2$, Fe, Ru all at high purity (99.9 at. % or better), in form of fine powders, mixed, pressed in pellets, sealed under vacuum in pyrex flask and heated up to 450°C for 15-20 hours. These first reaction products were then grinded, pressed in pellets again, sealed in quartz tubes and heated up to 1000°-1075°C for 50 hours. All these operations were carried out in a high purity argon atmosphere glove box (H$_2$O/O$_2$ less than 1 ppm). See Ref. 16 for more details.

Phase identification was performed by X-ray powder diffraction at room temperature (XRPD; PHILIPS PW3020; Bragg-Brentano geometry; CuK$_{\alpha 1, \alpha 2}$; range 15–120° $2\theta$ ; step 0.020° $2\theta$; sampling time 10 s). Rietveld refinement was carried out using the FULLPROF software;[17] by means of a LaB$_6$ XRPD standard an instrumental resolution file was obtained and applied during refinements in order to determine the micro-structural contribution to the XRPD peak shape. The diffraction lines were modeled by a Thompson-Cox-Hastings pseudo-Voigt function convoluted with an axial divergence asymmetry function and the background was fitted by a fifth-order polynomial. The following parameters were refined in the final cycle: the overall scale factor; the background (five parameters of the $5^{th}$ order polynomial), the zero offset in $2\theta$; the unit cell parameters; the specimen displacement; the reflection-profile asymmetry; the Wyckoff numbers not constrained by symmetry; the isotropic thermal parameters $B$; the anisotropic strain parameters.

All the analyzed Sm(Fe$_{1-x}$Ru$_x$)As(O$_{0.85}$F$_{0.15}$) samples crystallize in the tetragonal $P4/nmm$ space group at room temperature. Small amounts of SmOF (5-6 %) are present whatever the amount of Ru substitution. Besides SmOF, large amounts of additional secondary phases are present in SmRuAs(O$_{0.85}$F$_{0.15}$) ($x$=1.00), which prevents an accurate structural refinement and only the cell parameters can be reliably determined (see Table II).

As shown in Fig. 1, the cell parameter $a$ increases almost linearly with increasing Ru content; conversely the cell parameter $c$ exhibits an almost constant value up to 10% of Ru substitution and then a pronounced decrease takes place. The same behavior of the structural parameters was experimentally observed in both 1111 [14,15] and 122 [11,12,13] families. The increase of $a$ can be ascribed to the increased average size of the transition metal (*TM*) site within the structure, Ru being larger than Fe. Conversely the decrease of $c$ is mainly due to the decrease of the (*TM*)As layer thickness (see Fig. 2 and 3), determined by the increase of the As-*TM*-As bond angle (both arsenics lie in the same plane). In Table I the main crystallographic data of the samples under test are summarized.

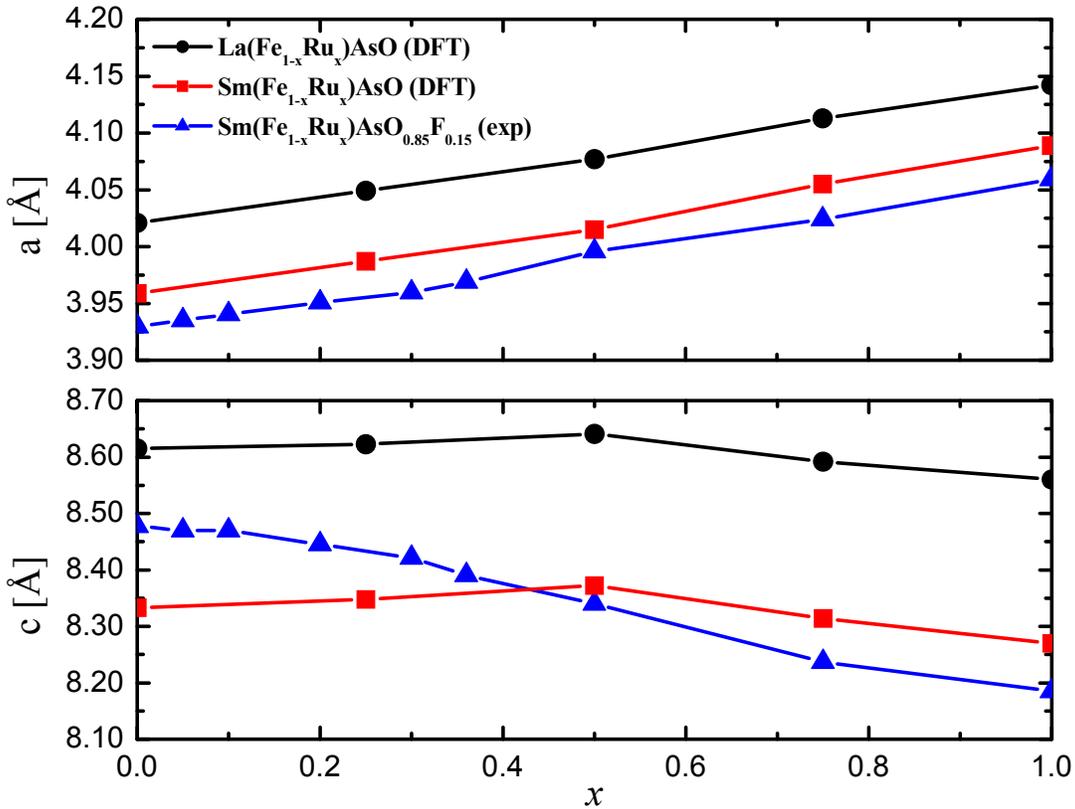

**Figure 1**: (Color online) dependence of the cell parameters of Sm(Fe$_{1-x}$Ru$_x$)As(O$_{0.85}$F$_{0.15}$) on Ru content $x$ (▲). The calculated lattice constants for La(Fe$_{1−x}$Ru$_x$)AsO (●) and Sm(Fe$_{1−x}$Ru$_x$)AsO (■) are plotted as comparison.

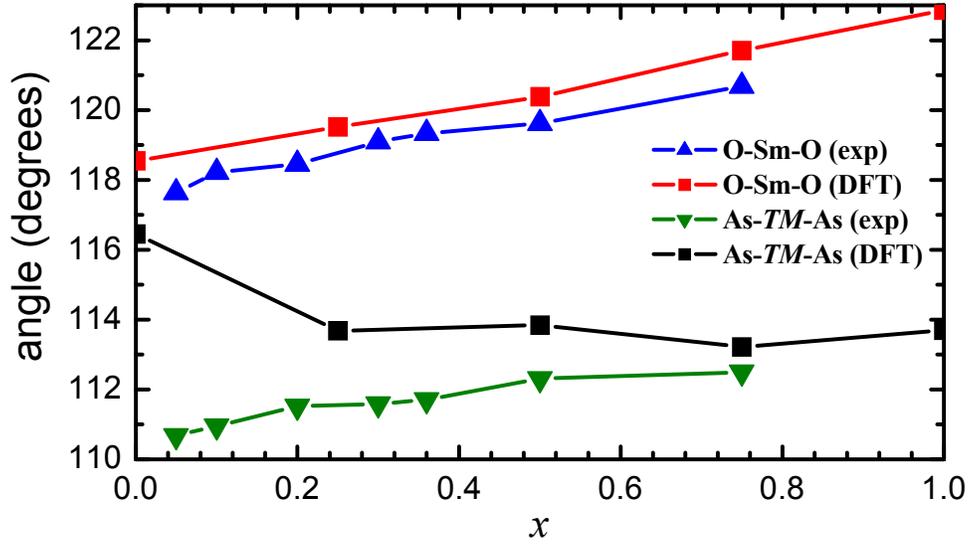

**Figure 2**: (Color Online) dependence of selected bond angles as a function of Ru content in Sm(Fe$_{1-x}$Ru$_x$)As(O$_{0.85}$F$_{0.15}$) (exp) and Sm(Fe$_{1-x}$Ru$_x$)AsO (DFT) respectively.

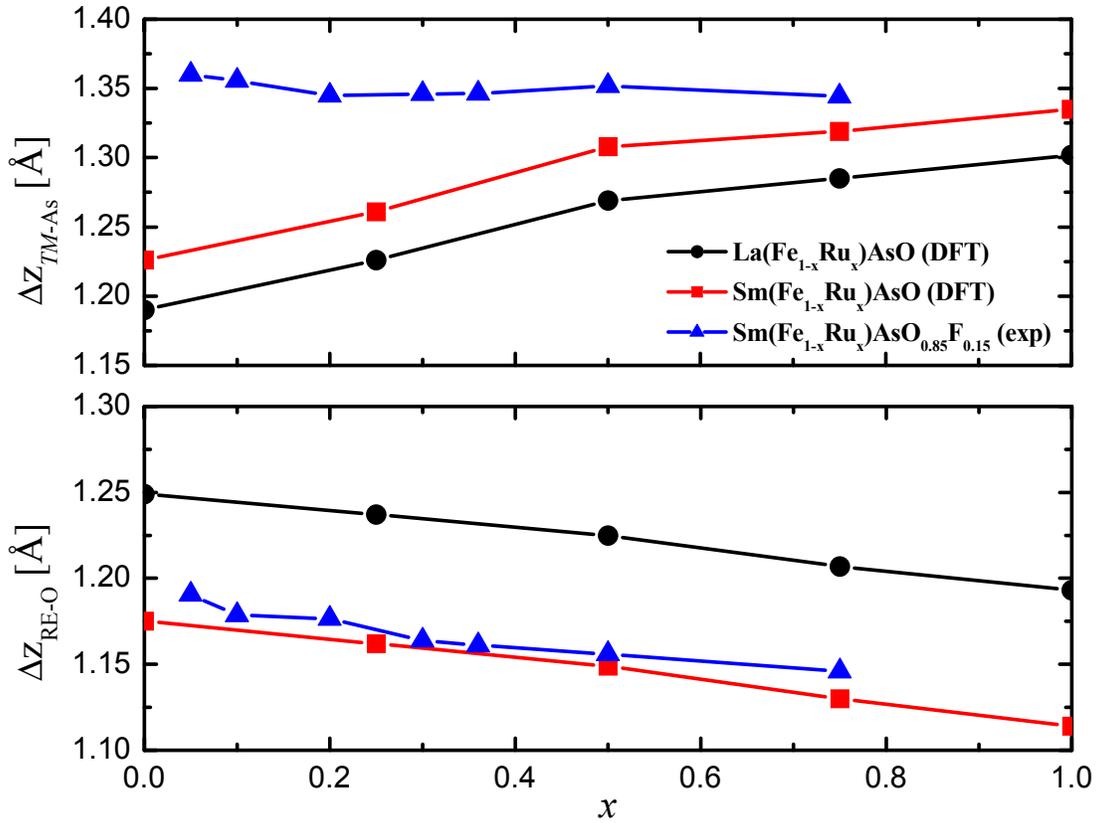

**Figure 3** (Color online). Calculated distances between the *TM*- and As-plane as well as the *RE*- and O-plane (*RE*=La,Sm); the corresponding experimental values for Sm(Fe$_{1-x}$Ru$_x$)As(O$_{0.85}$F$_{0.15}$) are plotted for comparison (▲). (●) refers to La(Fe$_{1-x}$Ru$_x$)AsO, (■) to Sm(Fe$_{1-x}$Ru$_x$)AsO respectively.

**Table I**: structural data for $Sm(Fe_{1-x}Ru_x)As(O_{0.85}F_{0.15})$ samples obtained by Rietveld refinement of XRPD data ($P4/nmm$ space group; origin choice 2).

|            | $x = 0.05$ | $x = 0.10$ | $x = 0.20$ | $x = 0.30$ | $x = 0.36$ | $x = 0.50$ | $x = 0.75$ |
|------------|------------|------------|------------|------------|------------|------------|------------|
| $a$ (Å)    | 3.9351(1)  | 3.9401(1)  | 3.9507(1)  | 3.9597(1)  | 3.9691(1)  | 3.9958(1)  | 4.0239(1)  |
| $c$ (Å)    | 8.4700(2)  | 8.4702(2)  | 8.4457(2)  | 8.4211(2)  | 8.3905(2)  | 8.3399(2)  | 8.2366(2)  |
| $z$ Sm     | 0.1406(1)  | 0.1391(1)  | 0.1393(2)  | 0.1382(2)  | 0.1384(2)  | 0.1386(2)  | 0.1391(2)  |
| $z$ As     | 0.6606(3)  | 0.6600(3)  | 0.6592(3)  | 0.6598(4)  | 0.6604(4)  | 0.6621(4)  | 0.6632(4)  |
| $R_f$ (%)  | 4.07       | 4.54       | 4.90       | 5.74       | 4.77       | 4.75       | 8.66       |
| $R_B$ (%)  | 6.92       | 7.47       | 6.88       | 8.44       | 6.77       | 6.43       | 9.21       |

### 3. Calculated structural, electronic and magnetic properties

In order to provide a theoretical framework to better understand the measured transport, magnetic and superconducting properties, we also investigated $RE(Fe_{1-x}Ru_x)AsO$ ($RE$=La,Sm) by first-principles calculations in the Density Functional Theory framework as implemented in the VASP package[18]. The Perdew-Wang [19] version of generalized gradient approximation (GGA) is used for the structural optimizations, and the local density approximation (LDA) (Perdew and Zunger [20]) for electronic and magnetic properties. A plane wave basis with a 353 eV cutoff is used and the electron-ion interaction is accounted by the projector augmented wave approach [21].

We performed our calculations both on $Sm(Fe_{1-x}Ru_x)AsO$ and on the prototype $La(Fe_{1-x}Ru_x)AsO$ structures for several Ru concentrations whereas F doping was considered in a rigid band model. Instead, the electronic and magnetic properties were calculated referring to La rather than Sm system, since the Sm-partially filled $f$ states are hard to treat within density functional theory. Actually, the substitutions in the iron-arsenide layers are the crucial ones in terms of transport properties, although different rare earth element does change structural properties and the $T_c$ [22].

We represent $La(Fe_{1-x}Ru_x)AsO$ system as ordered in superlattice structures within the periodic boundary conditions. Several Ru contents were considered as $x$= 0, 0.25, 0.5, 0.75, 1, which can be represented within a $\sqrt{2}\times\sqrt{2}\times1$ supercell, encompassing four formula units, as shown in Fig. 4. While for $x$=0.25 and 0.75 all of the possible Ru positions are equivalent, for $x$=0.5 two ordered structures are possible within the same given supercell: Ru and Fe atoms form a checkerboard structure in the former, or alternating stripes along the nearest neighbors in the latter.

Here we present the results for the checkerboard because more representative of a disordered structure with respect to the stripe arrangement. Since magnetic properties in LaFeAsO and related compounds are strongly dependent on the structure of the system we adopt the strategy outlined in Ref. 23 : we used the generalized gradient approximation (GGA) for the structural optimization of the supercell geometry and the atomic positions, the local density functional (LDA) for the magnetic properties. This choice improves the agreement between theoretical and experimental magnetic properties. In order to keep the computational effort under control, we performed all the structural minimizations in the non-magnetic state.

*(a) Structural properties*

The calculated values of the structural parameters are shown in Table II and compared with experimental results in Fig. 1,2 and 3. Our calculations show that $a$ increases upon Ru substitution, by about 3% going from $x$=0 to $x$=1. The agreement with experiment is excellent in terms of slope versus Ru content as it is the case of both La and Sm based systems. As for the absolute values, the calculations for $Sm(Fe_{1-x}Ru_x)AsO$ show a deviation from experiment well within 1% to be considered a pretty good agreement.

**Table II**. Calculated structural parameters for the La(Fe$_{1-x}$Ru$_x$)AsO and Sm(Fe$_{1-x}$Ru$_x$)AsO ordered compounds versus Ru concentration $x$: the lattice parameters $a$ and $c$ (in Å); the As coordinates in units of $c$; the Fe-As bond length (in Å); the distance between *TM* and As planes, $\Delta z_{TM-As}$, and rare earth to oxygen planes, $\Delta z_{RE-O}$, (both in Å); the Fe magnetic moments (in Bohr magnetons).

| $x$ for La(Fe$_{1-x}$Ru$_x$)AsO | a (Å) | c (Å) | $z_{As}$ | $z_{La}$ | $d_{Fe-As}$ (Å) | $\Delta z_{TM-As}$ (Å) | $\Delta z_{La-O}$ (Å) | O-La-O (degrees) | $\mu_{Fe}$ |
|---|---|---|---|---|---|---|---|---|---|
| 0.00 | 4.021 | 8.615 | 0.6382 | 0.1450 | 2.337 | 1.190 | 1.249 | 116.29 | 0.66 |
| 0.25 | 4.049 | 8.623 | 0.6423 | 0.1435 | 2.365 | 1.226 | 1.237 | 117.09 | 0.87 |
| 0.50 | 4.077 | 8.641 | 0.6469 | 0.1417 | 2.401 | 1.269 | 1.225 | 118.0 | - |
| 0.75 | 4.113 | 8.592 | 0.6495 | 0.1405 | 2.403 | 1.285 | 1.207 | 119.15 | - |
| 1.00 | 4.142 | 8.560 | 0.6521 | 0.1394 | - | 1.302 | 1.193 | 120.1 | - |
| $x$ for Sm(Fe$_{1-x}$Ru$_x$)AsO | a (Å) | c (Å) | $z_{As}$ | $z_{Sm}$ | $d_{Fe-As}$ (Å) | $\Delta z_{TM-As}$ (Å) | $\Delta z_{Sm-O}$ (Å) | O-Sm-O (degrees) | $\mu_{Fe}$ |
| 0.00 | 3.959 | 8.333 | 0.6471 | 0.1411 | 2.326 | 1.226 | 1.175 | 118.55 | 0.60 |
| 0.25 | 3.987 | 8.348 | 0.6511 | 0.1391 | 2.360 | 1.261 | 1.162 | 119.52 | 0.84 |
| 0.50 | 4.015 | 8.372 | 0.6562 | 0.1372 | 2.397 | 1.308 | 1.149 | 120.38 | - |
| 0.75 | 4.055 | 8.314 | 0.6586 | 0.1361 | 2.399 | 1.319 | 1.130 | 121.7 | - |
| 1.00 | 4.089 | 8.270 | 0.6614 | 0.1346 | - | 1.335 | 1.114 | 122.86 | - |

The calculated behavior of $c$, on the other hands, is more intriguing. In both La and Sm compounds, $c$ initially increases very slightly (up to 0.5 % at $x$=0.5) and it shows a small decrease (~ 1 %) for large $x$. Looking more carefully, we notice that both the quota of As atoms ($z_{As}$) and the distance between Fe and As planes, $\Delta z_{Fe-As}$ increase almost monotonically as a function of $x$ as shown in Fig. 3. The easiest explanation for the decrease of $c$, then, is that the larger size of Ru atoms leads to a larger $a$ value, and thereby to a tensile strain of the La(Sm)-O blocks. This strain is compensated by a corresponding decrease of the $z$-axis width of the latter block, and also of the distance among Fe-As and La-O blocks. The behavior of La and Sm based compounds follows parallel curves, with the smaller size of Sm determining almost always shorter distances and lattice parameters. Interestingly, however, this is not true for the inter-planar distance $\Delta z_{Fe-As}$, where we find a smaller value in the La case. A closer look at the structural properties, however, shows that the Fe-As interatomic distance is nearly the same, at equal composition, in the Sm and La compounds. Obviously, the latter result is due to larger As-Fe-As angles, made possible by the larger $a$ lattice parameter, and accompanied by in-plane As displacements. This finding is confirmed by the values of the O-Sm-O and As-*TM*-As bond angles reported in Fig. 2. Our calculations describe correctly the increase of the angle due to the tensile strain on the Sm-O block that flattens for higher Ru concentrations.

While the comparison between theory and experiment is satisfactory for the in-plane lattice constant, the same is not true for $c(x)$. To understand this point, we examine in Fig. 3 the difference between the quotas of the Fe and As planes as well as the one between the Sm(La) and O planes ($\Delta z_{Fe-As}$, $\Delta z_{Sm-O}$, and $\Delta z_{La-O}$). We immediately see that the behavior of $\Delta z_{Sm-O}$ is in pretty good agreement with experiment; also, $\Delta z_{Sm-O}$ is smaller than the corresponding distance $\Delta z_{La-O}$, consistent with the smaller Sm atomic size. However, the most intriguing result is found by comparing theory and experiment for $\Delta z_{Fe-As}$ in the Sm systems: we notice a good agreement in the Ru-rich compound ($x$=0.75), and a disagreement (around 0.1 Å, similar to that reported by many groups in the La case) for the Fe only compound ($x$=0). We can explain this finding by considering that our calculations are performed in a non-magnetic state: this is correct for the $x$=1 compound but the same does not hold for the $x$=0 compound. Indeed a good agreement for the As-*TM*-As bond angle exists only for high Ru content, while in the Fe rich phase the calculated value is overestimated. The influence of magnetism on the internal structural parameters has been extensively analyzed by Mazin and Johannes [24]; they recovered the agreement between theory and experiment by performing a spin-polarized calculation (in the correct magnetic phase), and argued that magnetism persists, on a local scale, even when doping suppresses the spin-density wave. Our findings are perfectly consistent with this point of view.

*(b) Magnetic and electronic properties*

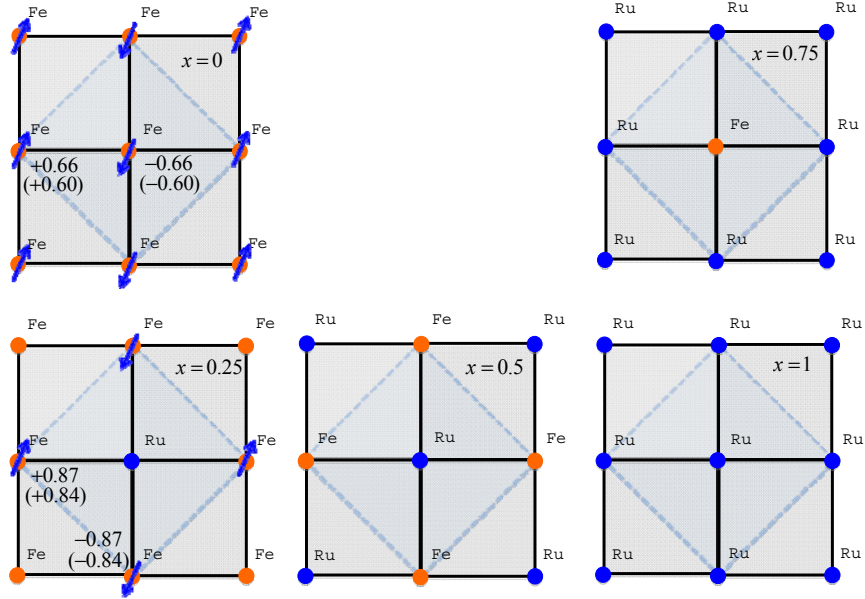

**Figure4**. (Color online) Structure and magnetic moment of $Sm(Fe_{1-x}Ru_x)AsO$ and $La(Fe_{1-x}Ru_x)AsO$ compounds shown along the basal *ab* plane. Moment values in parenthesis refer to $Sm(Fe_{1-x}Ru_x)AsO$

In the following paragraphs we will address the variations of the electronic and magnetic properties of $La(Fe_{1-x}Ru_x)AsO$ induced by Ru substitution, both directly and through the variations of structural properties. The dependence of the Fe-As distance upon structural ordering may in fact lead to some effect on the superconducting properties, which are strictly interconnected with magnetic properties, significantly dependent on the interatomic distances. In Table II we also show the magnetic moment of Fe ($\mu_{Fe}$) atoms in the anti-ferromagnetic (AFM) ordering for $x=0$ and 0.25. Test performed for higher Ru concentrations show that magnetism is on the verge of quenching, indeed Fe magnetic moment is found strongly dependent on the details of the calculations. We defer calculations at high Ru contents to future work.

Our calculations show that Ru atoms do not show any tendency to sustain a magnetic moment regardless their concentrations. This is a consequence of the different atomic size (thereby larger bandwidths), which does not allow any magnetic configuration of Ru in the given unit cell. As we will see later, the Ru *4d* band is in fact larger than the strongly peaked Fe *3d* band; as a consequence, magnetism is suppressed in $x=1$ compound. Clearly, this is not the case for $x=0$, where we used, the stripe phase, with ferromagnetic coupling between nearest neighbors, which is actually the ground state phase both experimentally and theoretically. For the case of $x=0.25$, the situation is more complex and we need to make reference to Fig. 4. Within the unit cell considered here, symmetry considerations rule out an AFM alignment of chains parallel to Cartesian axes. Assuming an AFM alignment within diagonal chains, the third Fe atom of the unit cell is magnetically frustrated and has a nearly vanishing moment.

As for the magnitude of the magnetic moment, our calculations show that the Fe moment significantly increase from $x=0$ to $x=0.25$, consistent with the larger Fe-As distance. The magnetic solutions is stable over the non-magnetic one, in our calculations, up to $x=0.5$, where magnetic and non magnetic solution have very similar energies. This seems to indicate that $x=0.5$ is the threshold for the stability of magnetism.

In order to investigate further the changes introduced in the electronic structure by Ru substitution, in Fig. 5 we show the electronic density of states (DOS) of La(Fe$_{1-x}$Ru$_x$)AsO computed for different Ru contents, in the non magnetic state. At first sight, the general shape of the DOS looks reasonably similar in all compounds, with the exception of the pure Ru ($x$=1) case. In particular, the DOS around the Fermi level have very similar shapes (with a large peak from Fe *3d* states at ~ -0.5 eV). A closer look at Fig 5 shows however a systematic trend to a larger width of the wide peak just below $E_F$. Moreover, the DOS at the Fermi level (N($E_F$)), decreases upon Ru substitution.

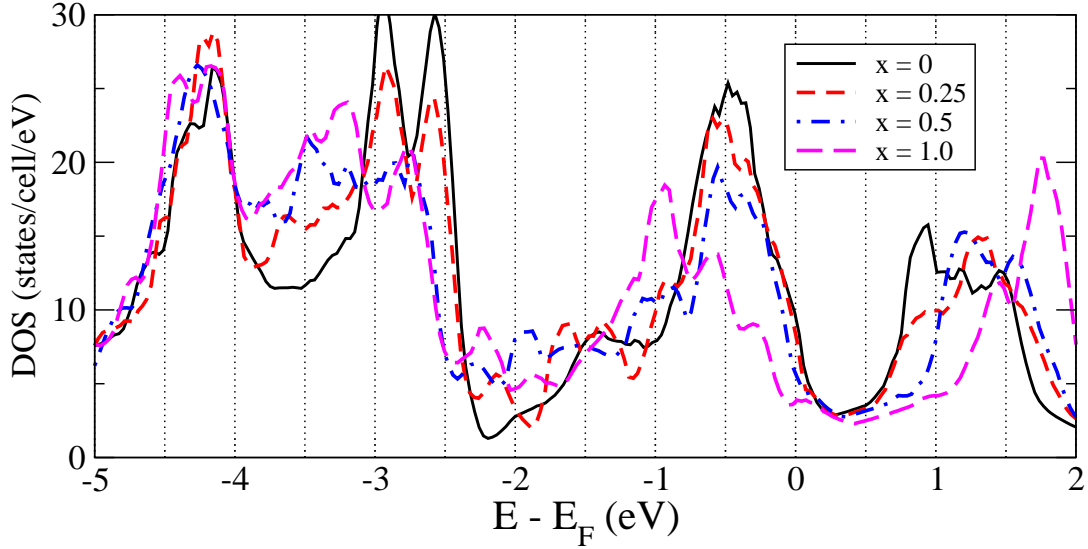

**Figure 5**: (Color online) total DOS for the La(Fe$_{1-x}$Ru$_x$)AsO at different Ru concentrations.

A more detailed view is given in Fig. 6 by the study of the DOS projected (PDOS) in the Fe and Ru atomic-spheres. We see important differences between the Fe and Ru PDOS: in the former, in fact, we see the large peak at ~ -0.5 eV mentioned previously, which is largely broadened in the Ru PDOS. This broadening of Ru derived bands of course, correlates with the different magnetic behavior of the two atoms. In the same way, we can see that the contribution of Ru is larger in the low energy region (below 2.5 eV), again with broader bands that is the sign of a larger hybridization with As states.

The Fe PDOS at the different compositions do not show dramatic differences. In particular, they do not show any visible offset relative to $E_F$, which would derive from their larger or smaller filling. This implies little (global) charge doping even in the presence of a very large compositional doping. Fig. 5 and 6 are in good agreement with calculations on Pr(Fe$_{1-x}$Ru$_x$)AsO [14] showing similar trends in the evolution of DOS and its Fe and Ru component with Ru content. The same pronounced reduction of the DOS at $E_F$ is found in both compounds, DOS projected on Ru is broader than that on Fe and a similar Ru-As hybridization feature shows up 3.5 eV below $E_F$, suggesting strong similarity in the effect of Ru substitution in electronic structure of 1111 family compounds.

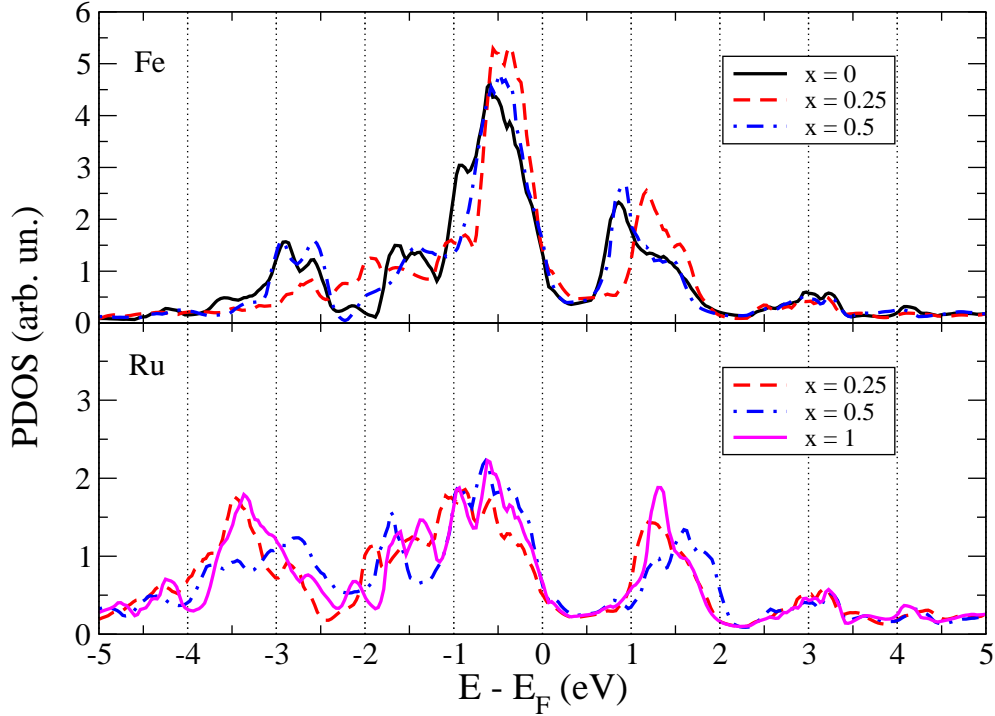

**Figure 6**: (Color online). Fe and Ru *d*-states contribution to the PDOS inside an atomic sphere of 1.1 Å for La (Fe$_{1-x}$Ru$_x$)AsO system. The upper panel shows the PDOS on Fe and the lower one on Ru.

In Fig. 7 we plot the band structures close to E$_F$ for the different compositions. Since we use a doubled unit cell relative to the crystallographic one (four against two formula units in the crystallographic cell), the *M* point of the latter folds into the Γ point of our supercell. Therefore, both holes and electron bands appear around our Γ point. We notice immediately that, although the global centre of gravity of bands does not change with Ru content (no extra charge is induced by the isovalent substitution), the *e* and *h* bands shift in such a way that both the number of *e* and *h* increase with *x*. Remarkably, however, the shape of the bands changes in such a way (band dispersions increase with *x*) as to lead only to small changes of the Fermi surfaces. We also notice an offset in the $d_{z^2}$, $d_{xy}$ bands just below E$_F$, which goes up in energy with increasing *x*.

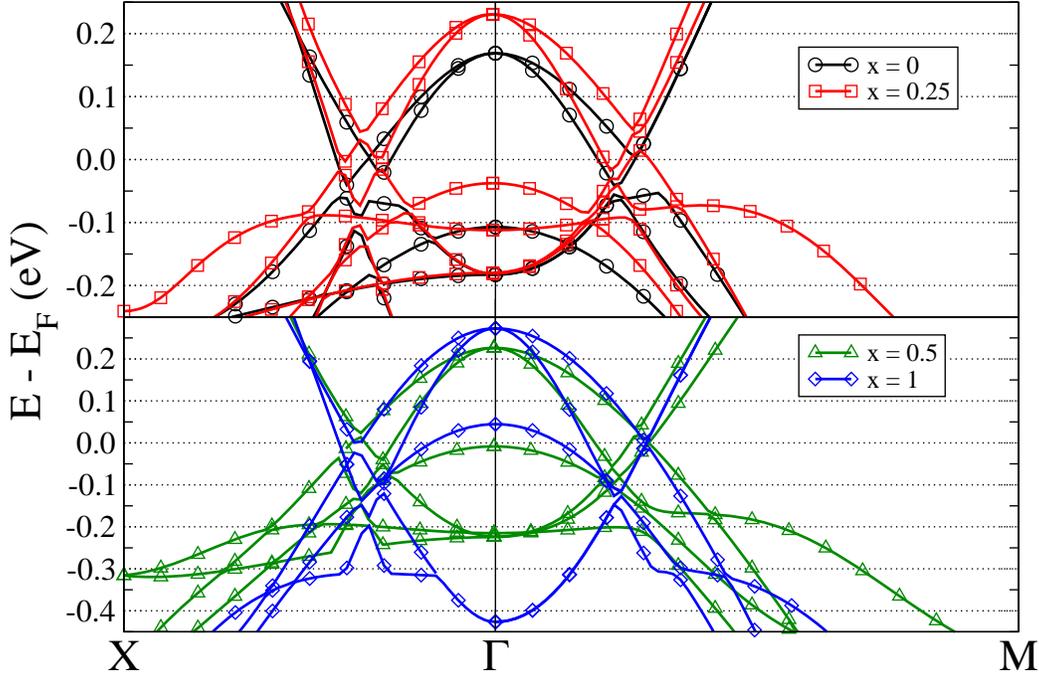

**Figure 7**: (Color online) Energy bands (in the non-magnetic configuration) for La(Fe$_{1-x}$Ru$_x$)AsO at different Ru concentration. Most bands not crossing the Fermi energy are not shown for sake of clarity.

In order to get more insights on the doping induced by Ru substitution, we computed the Hall coefficients $R_{\alpha\beta\gamma}$ at the end-point compounds ($\alpha\beta\gamma$ being the Cartesian components). The reason for this choice is that in the supercell approach the folding of bands with the resulting crossings right around $E_F$ makes a reliable estimate of $R_{\alpha\beta\gamma}$ nearly impossible. Our approach is based on Bloch-Boltzmann theory and gives $R_{\alpha\beta\gamma} = E_\beta / j_\alpha B_\gamma = \sigma_{\alpha\beta\gamma} / \sigma_{\alpha\alpha}\sigma_{\beta\beta}$, where:

$$\sigma_{\alpha\beta} = \frac{e^2\tau}{\Omega} \sum_{n,\mathbf{k}} v_\alpha(n\mathbf{k}) v_\beta(n\mathbf{k}) \left(-\frac{\partial f}{\partial \varepsilon_{n\mathbf{k}}}\right) \quad (1)$$

$$\sigma_{\alpha\beta\gamma} = -\frac{e^3\tau^2}{\hbar\Omega} \sum_{n,\mathbf{k}} v_\alpha(n\mathbf{k}) \left[\mathbf{v}(n\mathbf{k}) \times \nabla_\mathbf{k}\right]_\gamma v_\beta(n\mathbf{k}) \left(-\frac{\partial f}{\partial \varepsilon_{n\mathbf{k}}}\right). \quad (2)$$

Here, $f$ is the Fermi-Dirac function, $\Omega$ is the normalization volume, and $\varepsilon_{n\mathbf{k}}$ is the energy of band $n$ at point $\mathbf{k}$ in the Brillouin zone. Each of the two tensors above are obtained by a sum over bands. In order to investigate the contribution of holes ($h$) and electrons ($e$), we consider $R_{\alpha\beta\gamma}$ into two distinct $e$ and $h$ terms, $R_{\alpha\beta\gamma}^{(h,e)} = \sigma_{\alpha\beta\gamma}^{(h,e)} / [\sigma_{\alpha\alpha}\sigma_{\beta\beta}]^{(total)}$. In this way we obtain contributions summing up to the total value. This does not correspond, of course, to the Hall coefficients, which would result from hypothetical compounds containing $e$ or $h$ only. Furthermore, since the experimental measurements are performed on polycrystalline samples, we average over the three independent tensor components ($R_{xyz} \neq R_{yzx} = R_{zxy}$) to obtain $R_H$.

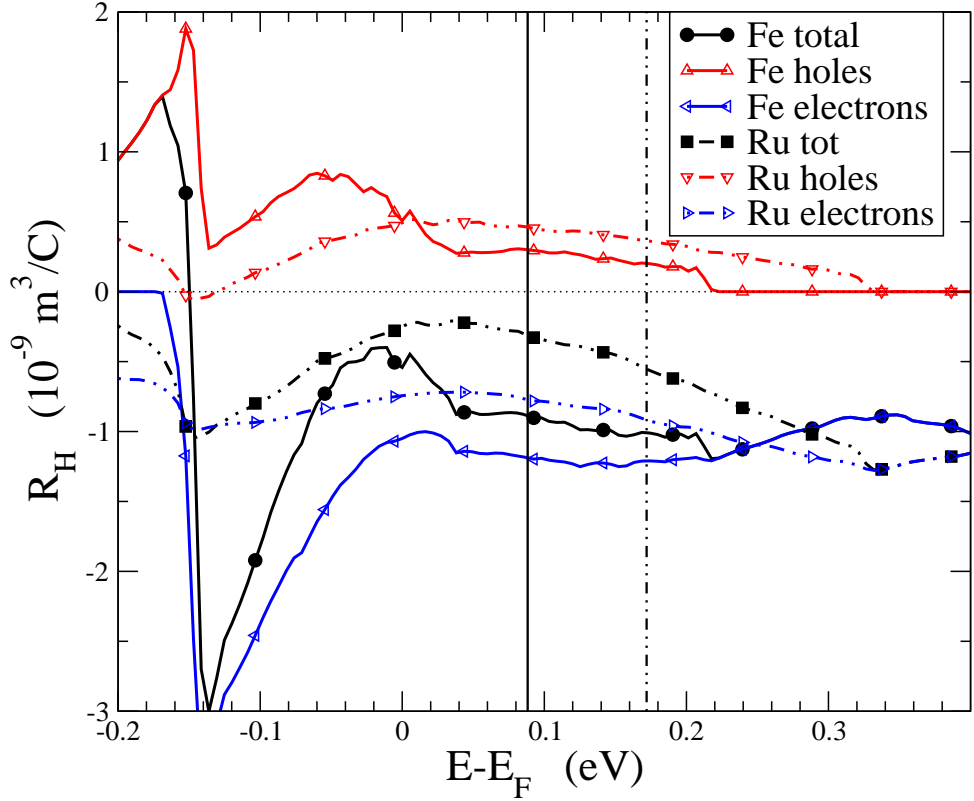

**Figure 8**: (Color online) Hall coefficients for LaFeAs(O$_{0.85}$F$_{0.15}$) (full lines) and LaRuAs(O$_{0.85}$F$_{0.15}$) (broken lines) as divided into hole and electron contributions (see text for further details).

In Fig. 8 we plot $R_H$ for LaFeAs(O$_{0.85}$F$_{0.15}$) and LaRuAs(O$_{0.85}$F$_{0.15}$) ($x=0$ and $x=1$ respectively). Fluorine doping is treated within a rigid band scheme. For each compound the vertical bars marks the Fermi energy corresponding to the experimental F doping within a rigid band model. We first notice a large compensation in both compounds; this is particularly true for LaRuAsO compound, where $e$ and $h$ sum up to a quite small $R_H$ value. In the Fe compound, however, the vicinity to the complete filling of the $h$ bands makes the compensation only partial, leading to a markedly $e$-like compound. In fact, the larger bandwidths found in the Ru compound are the most relevant difference, which yields to the smaller $R_H$. We also mention that wild variations appear as a function of the distance from the Fermi level, arising from to the appearance/disappearance of some bands at low energy, and to the Hall tensor components with magnetic field parallel to the $a$ and $b$ axes.

Finally, we note that our calculation and those of Ref. [25] on BaFe$_{1-x}$Ru$_x$As show that Ru substitution on 1111 and 122 compounds has similar effects on the electronic structure: it does not induce new bands or alter the charge balance between electron and hole densities, instead it results in a broadening of the $d$-bands, a bit more pronounced in our case, with a lessening of the DOS at E$_F$.

## 4. Normal state properties

*(a) Resistivity*

Having examined how the electronic structure near $E_F$ evolves as a function of Ru substitution, we now turn back to experimental results. In Fig. 9 we present the temperature dependence of electrical resistivity ρ for samples with $x \leq 0.32$ (left panel) and for samples with $x > 0.33$ (right panel). A general trend is observed, even if a strict dependence of ρ with $x$ is not present: the resistivity ρ(T) increases for $x<0.25$, it does not change significantly in the range $0.30<x<0.40$ and it progressively decreases for further Ru substitution. Substantially the same behavior has been reported for $Nd(Fe_{1-x}Ru_x)As(O_{0.89}F_{0.11})$[15], $Nd(Fe_{1-x}Co_x)As(O_{0.89}F_{0.11})$[10] and $La(Fe_{1-x}Co_x)As(O_{0.89}F_{0.11})$[10] compounds with Ru and Co substitutions at Fe site.

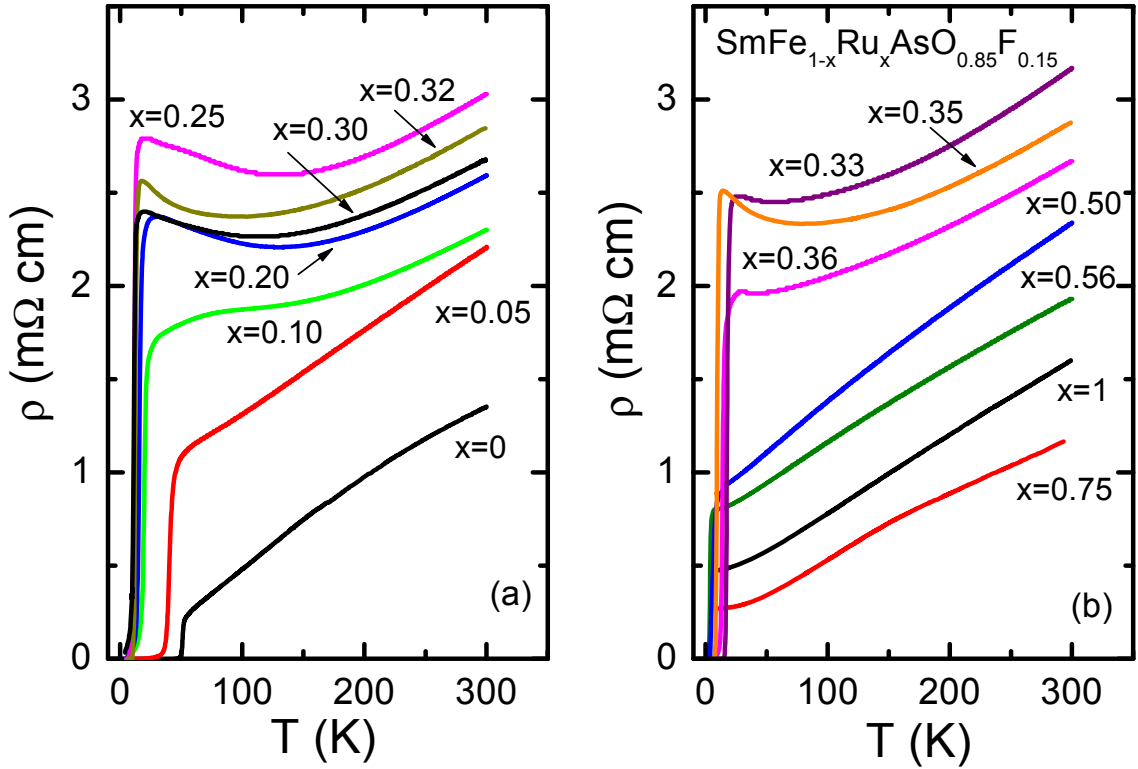

**Figure 9**: (Color online) resistivity vs T for $Sm(Fe_{1-x}Ru_x)As(O_{0.85}F_{0.15})$ samples.

Above 200 K all curves are roughly parallel and increase linearly with temperature. With decreasing temperature an upturn emerges in the curves with $x$ in the range 0.05-0.36 that are characterized by resistivity values exceeding 1 mΩ cm around 50 K . Similar upturns have been often observed in 1111 compounds, in which disorder has been introduced with substitution at Fe site[10,15,26] and by means of irradiation [27, 28], and in Fe(Te,Se)[29]. In NdFeAs(OF) irradiated with alpha particles[28] and in Fe(Te,Se)[29] the upturn is logarithmic and the magnetoresistance is negative that strongly suggests Kondo type scattering with magnetic impurities. In both these compounds it is believed that magnetic scattering by localized moments is due to Fe ions lying out of the Fe planes.
In 1111 compounds substituted with magnetic ions (Mn and Co) [10, 26] at Fe site, the nature of the upturn is still under debate because no simple correlation between the resistivity upturn and the amount of substitution has been found.

*(b) Hall Effect and magnetoresistivity*

In order to investigate different transport regimes as a function of Ru substitution, magnetoresitivity and Hall effect measurements were carried out.

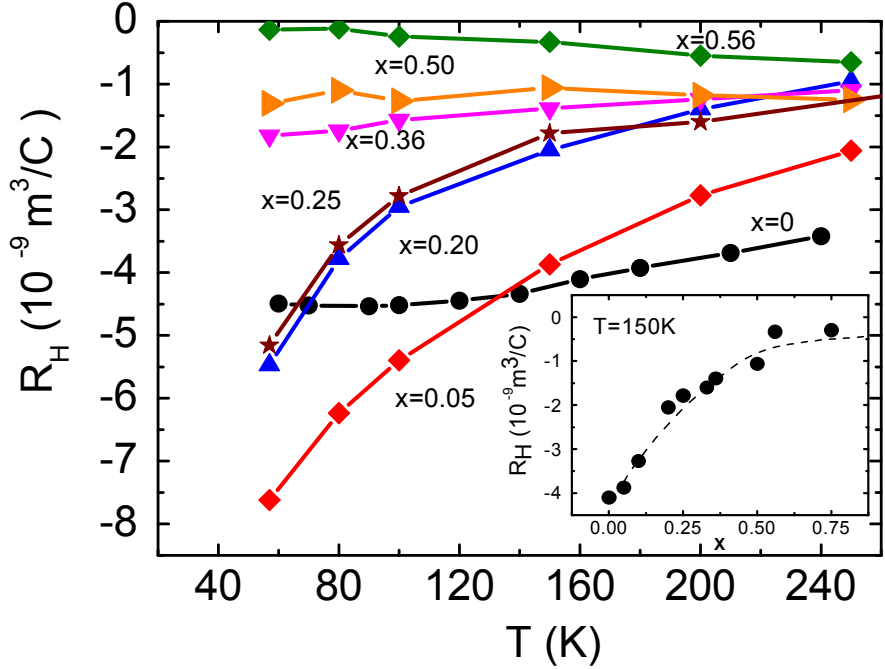

**Figure 10**: (Color online) $R_H$ as a function of T for selected $x$ values. Inset: $R_H$ as a function of $x$ at 150 K; the dashed line is a guide for the eyes.

Fig. 10 shows the Hall resistance $R_H$ as a function of temperature for $x$ = 0, 0.5, 0.20, 0.36, 0.5, 0.56. $R_H$ is negative for all the samples and exhibits steeper $T$ dependence at some selected $x$ values (0.05, 0.20, 0.25) which roughly correspond to the samples exhibiting an evident low temperature upturn. Again, similar behavior has been reported for $Nd(Fe_{1-x}Ru_x)As(O_{0.89}F_{0.11})$[15], $Nd(Fe_{1-x}Co_x)As(O_{0.89}F_{0.11})$[10] and $La(Fe_{1-x}Co_x)As(O_{0.89}F_{0.11})$[10].

Above a certain temperature (around 140 K) where low temperature upturns make a minor contribution the absolute values of $R_H$ decrease monotonically with increasing $x$, as shown in the inset of Fig. 10, where $R_H$ at 150 K is plotted as a function of $x$. The overall behavior could suggest that the dominant charge carriers are electrons at all the doping levels and their concentration increases with increasing Ru content. However, the ab-initio calculations depict the Ru rich phase as strongly compensated, which prevents from extracting reliable information on the actual carrier concentration from $R_H$. Indeed the strong reduction of $|R_H|$ with $x$ seems to be more related to the ongoing compensation than to an increase of the actual charge carrier densities. Keeping this in mind, we may be satisfied of the agreement between experiments and theory for the Ru compound, both pointing to a very small $|R_H|$. For the pure Fe case, on the other hand, the agreement is worse, with a quite larger experimental $|R_H|$. We believe that this can be a further warning that (on a local level) Fe keeps its magnetic polarization, leading to the a depletion of states around $E_F$ and to a correspondent increase of $|R_H|$. Test calculations supported the plausibility of this interpretation.

Magnetoresistivity in the normal state (T=57 K) was measured up to 9 T. $[\rho(B)-\rho(0)]/\rho(0) = \Delta\rho/\rho(0)$ is plotted in fig. 11 as function of $B^2$ for the samples with $x$ = 0, 0.05, 0.10, 0.20, 0.25, 0.36, 0.5, 0.56, 0.75. $\Delta\rho/\rho(0)$ is positive for all the Ru concentrations and this rules out that magnetic scattering is a significant mechanisms in these compounds. All the samples exhibit a roughly $B^2$ dependence. In a single band system the cyclotronic magnetoresistivity is due to the leading order by $\Delta\rho/\rho_0 \approx (B\mu)^2$. On the other hand, if the system has two bands of electrons and holes, whose mobilities and conductivities are $\mu_e$, $\mu_h$, $\sigma_e$ and $\sigma_h$, respectively, the cyclotron magnetoresistivity is given by $\Delta\rho/\rho_0 \approx \frac{\sigma_h \sigma_e}{(\sigma_h+\sigma_e)^2}(|\mu_h|+|\mu_e|)^2 B^2 \equiv (\mu_{MR} B)^2$ where $\mu_{MR}$ is an effective carrier mobility which is a good parameter to quantify the effect of disorder as a function of doping. $\mu_{MR}$ has been plotted in Fig.12.

The carriers mobility can be alternatively evaluated from Hall data as $\mu_H = |R_H|/\rho$, provided a single band description applies. As discussed above, this is clearly not the case mainly in the Ru rich phase. However, we also plot in Fig. 12 $\mu_H$ at 57 K. In general, mobilities extracted by these different techniques seldom match closely, even in single band systems because, a numerical coefficient of the order of unity and dependent on the scattering mechanism has to be taken into account to extract the cyclotron mobility $\mu_{MR}$ from the relationship $\Delta\rho/\rho_0 \approx (B\mu)^2$ [30]. In the present case we can see that $\mu_{MR}$ is 1-2 orders of magnitude larger than $\mu_H$ and assumes the largest value for $x=0$ and 0.75 whereas $\mu_H$, decreases substantially with $x$ as a consequence of the strong reduction of $|R_H|$. These remarkable differences in values and behavior of mobilities are due to multiband nature of these compounds. Hence, we assume that the mobility values inferred by magnetoresistance data are more reliable. Indeed, the carrier mean free path can be tentatively evaluated as $l = \mu \frac{m}{e} v_F$ where $m$ and $e$ are the electron mass and charge and $v_F=1.3\times10^5$ m/s is the Fermi velocity[31]. The obtained values reported in the right y axis of Fig. 12 are in the range 1-10 nm if calculated by $\mu_{MR}$ and vary from 1 nm to unreliably small values (0.1-0.01 nm) if calculated by $\mu_H$.

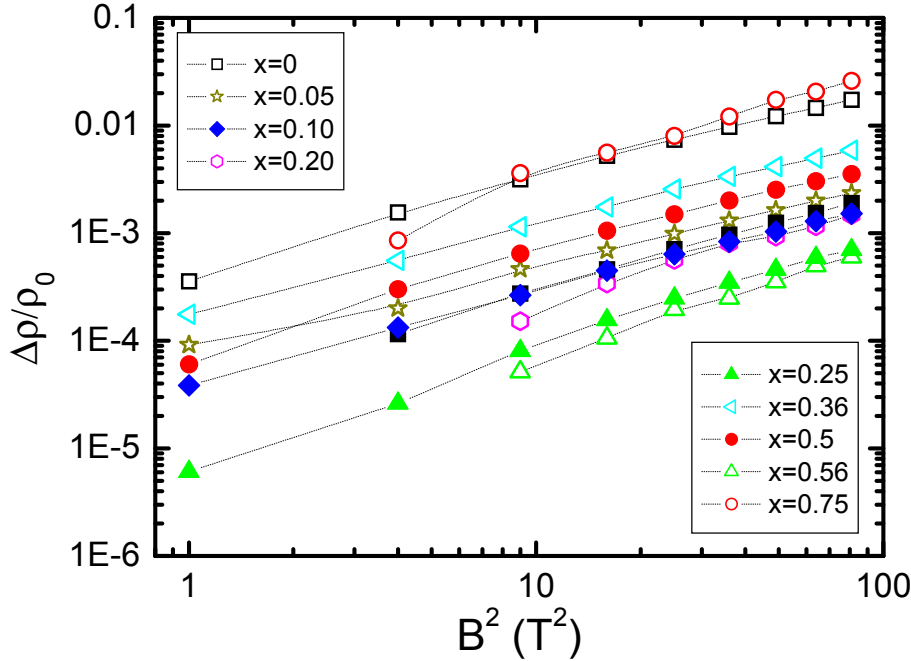

**Figure 11** (Color online): magnetoresistivity vs $B^2$ measured at 57 K for $x=$ 0, 0.05, 0.20, 0.36, 0.5, 0.56, 0.75.

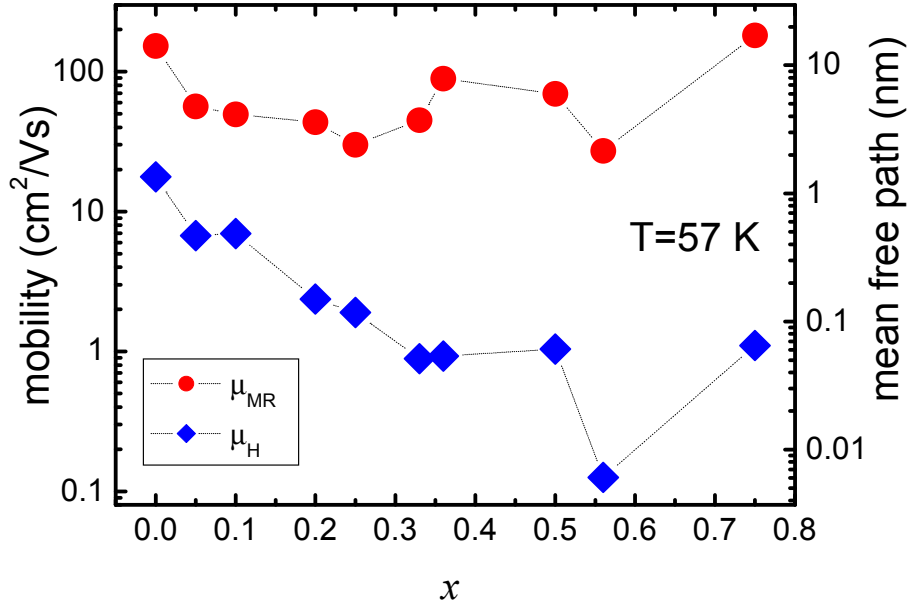

**Figure 12**: (Color online) $\mu_H$ evaluated by Hall effect (diamonds) evaluated and $\mu_{MR}$ evaluated by magnetoresistivity (squares) versus $x$. The mean free path is reported in the right y axis (see text for details).

*(c) Transport regimes of Fe rich and Ru rich phases*

The transport behavior of Fe-rich phase, as shown by the temperature dependence of resistivity and Hall effect, suggests that in this regime disorder plays a major role. At low Ru content ($0.05 \leq x \leq 0.36$) both $\rho$ and $R_H$ exhibit low temperature upturns (see Fig.s 9 and 10). Magnetic and non-magnetic origin of the upturn can be invoked: since magnetoresistivity measured in the temperature range where upturn emerges is positive at all Ru contents (see fig.3), we safely rule out that magnetic scattering can contribute to such upturns.

On the other hand, the upturn emerges for resistivity values above 1 mΩcm; if we regard this system as two-dimensional and we calculate the sheet resistance per FeAs layer, this resistivity value corresponds to the sheet resistance R =$\rho/c \approx$12 kΩ, which is of the same order of magnitude of the inverse minimum metallic conductivity ~ $e^2/h$ [32].

Considering a not negligible uncertainty on the resistivity value due to the polycrystalline nature of the samples under test, an exact correspondence is unlikely. Therefore we conclude that Anderson localization could better account for resistivity and Hall effect temperature dependence above $T_c$. We can conclude that in agreement with ab-initio calculation, in the Fe-rich phase Ru substitutions act as not magnetic defects, and strongly modify the conduction regime.

Different situation occurs in the Ru-rich phase ($x>0.5$) where resistivity values decrease with increasing Ru content and both $\rho$ and $R_H$ exhibit a metallic behavior. This is certainly due to the increasing bandwidth of Ru-rich compositions as compared to Fe-rich ones, as predicted by ab initio calculations.

**5. Superconducting state properties**

*(a) Upper critical Field*

Magnetoresistivity measurements of the Fe-rich samples were performed in high magnetic field at GHMFL, Grenoble, in order to evaluate the upper critical field. $\mu_0 H_{c2}$ was evaluated at the 90% of the resistive transition is plotted in Fig. 13 for $x$ = 0, 0.05, 0.10, 0.2, 0.3 and 0.36. The slopes -$\mu_0 dH_{c2}/dT$ are plotted as a function of $x$ in Fig. 14. The values are rather large for all the samples : -$\mu_0 dH_{c2}/dT$ is 11.3 T/K for the not substituted sample and then it decreases down to 5 T/K for $x$ =0.10 and fluctuates around this value.

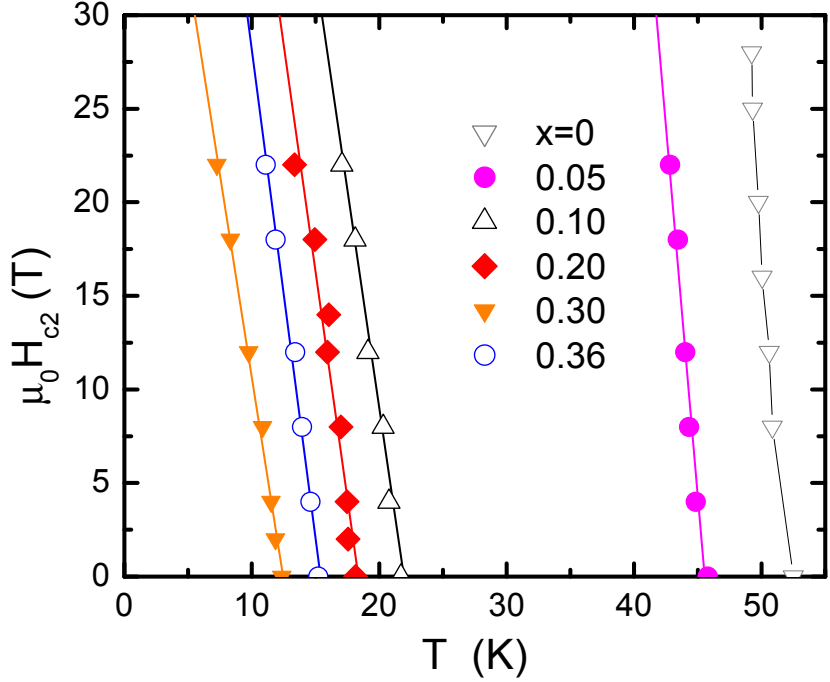

**Figure 13**: (Color online) $\mu_0 H_{c2}$ evaluated at the 90% of the resistive transition for $x$= 0, 0.05, 0.10, 0.2, 0.3, 0.36.

In a BCS framework the slope of $H_{c2}$ close $T_c$ is given by the following relationship:

$$\left|\frac{dH_{c2}}{dT}\right|_{T_c} \propto T_c\left(1+\frac{\xi_0}{l}\right) \quad (3)$$

where $\xi_0$ is the BCS coherence length, and $l$ is the electron mean free path. Thus, in the clean limit, ($\xi_0 < l$) $|dH_{c2}/dT_c|$ should scale only with $T_c$, whereas in the dirty limit, ($\xi_0 > l$), $|dH_{c2}/dT_c|$ is expected to increase with decreasing the mean free path. In the inset of Fig. 14 $\mu_0|dH_{c2}/dT_c|/T_c$ is plotted as a function of $T_c$. For comparison data relevant to $x$=0 and different F content samples are also shown[33]. It is interesting to note that the F doped samples identify a constant value of $\mu_0|dH_{c2}/dT_c|/T_c \sim 0.2$ T/K$^2$ that should represent the "clean limit" value. In fact, we can reasonably assume that F substitution which is out of the Fe-As layer slightly affects the mean free path. The Ru substituted samples with Tc lower than 20 K ($x \geq 0.2$) move progressively far from this value, exhibiting larger $\mu_0|dH_{c2}/dT_c|/T_c$ as expected in dirty limit. The mean free path as evaluated by $\mu_{MR}$ qualitatively supports this view: in the undoped sample $l$ is more than 10 nm which is larger than the coherence length value ($\approx$2-2.5 nm) estimated for a Sm-1111 sample with optimal $T_c$.[33] In the heavier doped samples $l$ drops down to 3 nm, whereas the coherence length which has been evaluated 4 nm for $T_c$=33 K,[33] is expected to increase progressively with decreasing $T_c$.

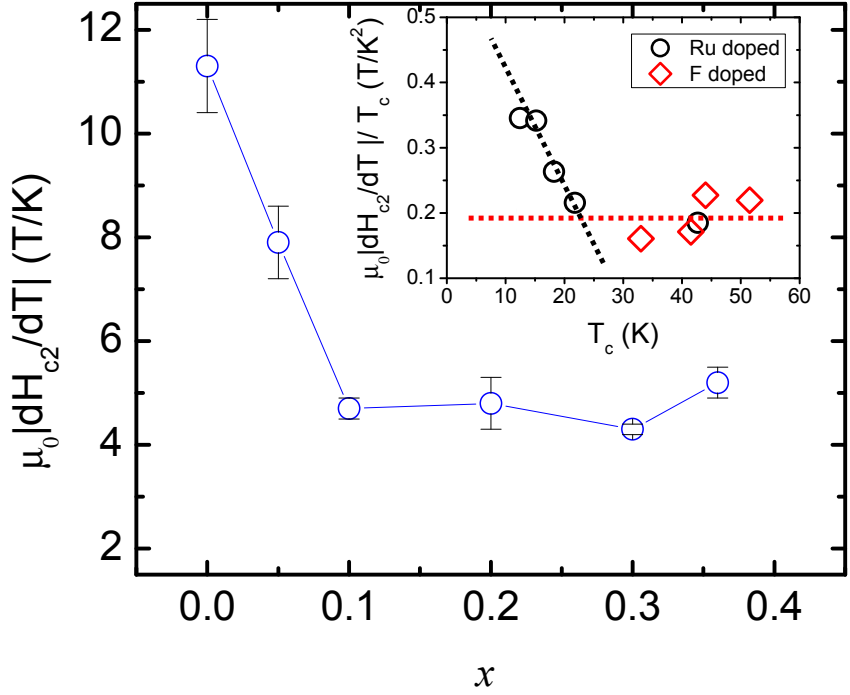

**Figure 14**: (Color online) The slopes $\mu_0|dH_{c2}/dT|$ are shown versus *x*. In the inset $\mu_0|dH_{c2}/dT_c|/T_c$ is plotted as a function of $T_c$. See text for details.

*(b) $T_c$ vs Ru-concentration behavior*

The critical temperature as shown in Fig. 15(a) progressively decreases with increasing Ru content. To explore a possible relationship between $T_c$ and the normal state resistivity, in Fig. 15 (b) ρ(0) is plotted as a function of *x*, where ρ(0) is the zero temperature linear extrapolation of ρ(T). Three different regions can be identified: with increasing *x* from 0 to 0.2, $T_c$ decreases from 51 K to 20 K and the resistivity steeply increases; for *x* between 0.2 and 0.4 $T_c$ values are scattered in the range 15-20 K and the resistivity is rather constant; beyond *x*=0.4 $T_c$ starts decreasing again and it is suppressed below 2 K for *x* =0.75, whereas the resistivity values progressively decrease.
In order to explain this behavior three different aspects should be considered: *i*) the role of doping, *ii*) the role of disorder, *iii*) the role of Fe magnetism.

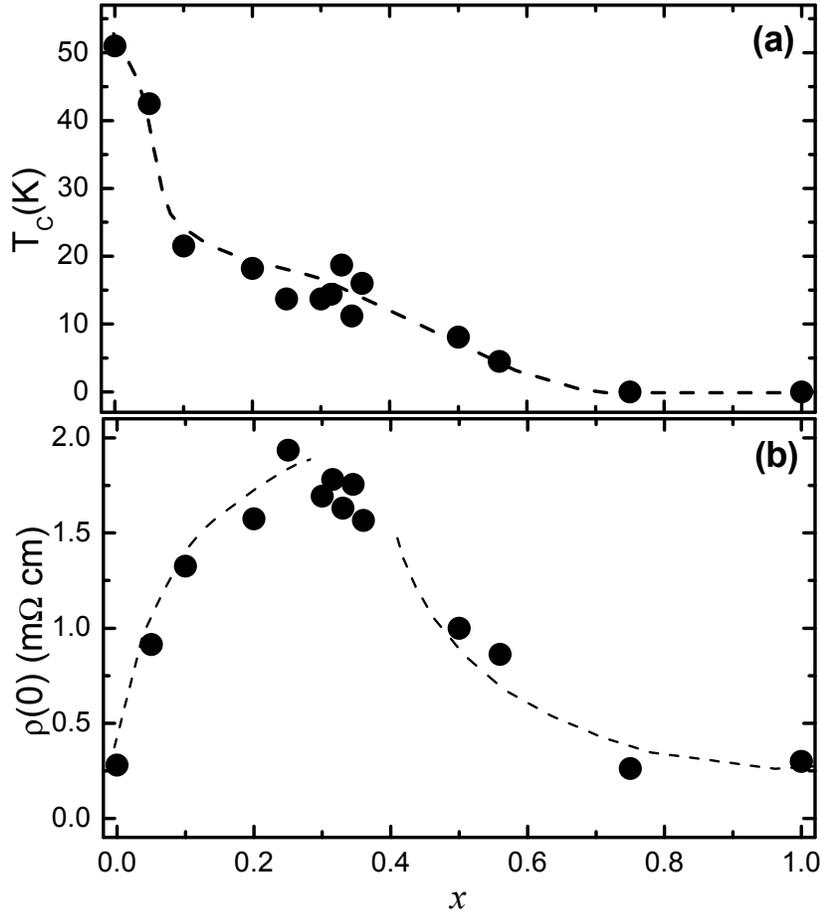

**Figure 15**: (a) Tc versus $x$ and (b) $\rho(0)$ vs $x$ for $Sm(Fe_{1-x}Ru_x)As(O_{0.85}F_{0.15})$ compounds where $\rho(0)$ is the linear extrapolation to zero temperature of the curves in the range 250-300 K. The lines are guides for the eyes.

*i) Role of doping*
To discuss the role of doping it is worth noticing that in this sample series superconductivity is strongly resilient to Fe substitution with Ru, if compared to the case of Fe substitution with Mn or Co. Indeed in $RE(Fe_{1-x}TM_x)As(O_{0.89}F_{0.11})$ (*RE*=La, Nd; *TM*=Co, Mn) compounds superconductivity is suppressed for Co amount of 7% and 11% and *RE* =La and Nd, respectively and for Mn amount of 1% and 4% and *RE* =La and Nd, respectively [10].
This different behavior can be well understood considering Mn and Co substitutions. In the case of Co, which shifts rigidly the Fermi level, electrons donated by Co atoms fill the hole pockets around the Γ point, causing the disappearance (or shrinkage) of the hole-Fermi-surfaces around Γ. Isoelectronic substitution of Fe with Ru instead, as shown by ab-initio calculation, preserves the existence of both *h* and *e* sheets. This evidence strongly suggests that one crucial point for occurrence of superconductivity is the existence of both *h* and *e* Fermi sheets. Since this feature is maintained by Ru substitution, changes in the electronic structure do not affect significantly the superconductivity and this gives the opportunity of investigating the role of disorder over a rather wide range of substitution.

*ii) Role of disorder*

Transport data suggest that the Fe-rich phase is strongly affected by disorder. Carrier mobility ($\mu_{MR}$) follows a non monotonic behavior as a function of *x*: it drops by a factor 3 upon substitution of a small amount of Ru, then it progressively decreases before reaching again larger mobility values for *x* around 0.5. As for the initial sharp decrease and the following smoother decrease, this behavior just mirrors the suppression of $T_c$ which is steep up to *x* =0.2 and then it fluctuates in the range 15-20 K for *x* in the range 0.2 to 0.4. Also the upper critical field undergoes a crossover from clean to dirty limits with increasing *x* above 0.2.

Interestingly our data show a steeper $T_c$ decrease as a function of Ru in comparison with data in ref.15. Accordingly $\rho(T)$ and $|R_H(T)|$ behaviors (low temperature upturn) indicate that disorder is more important in our samples.

All these are clear evidences of the role of disorder in tuning superconducting properties in pnictides superconductors.

The effect of disorder on $T_c$ has been largely discussed theoretically.[3] Strong suppression due to interband impurity scattering is expected within the $s^\pm$ model while conventional s-wave pairings are not strongly affected by disorder. In order to make a quantitative comparison the reduced scattering rate $g = \hbar\Gamma/2\pi k_B T_{c0}$ is the key parameter to be considered ($\Gamma$ is the scattering rate and $T_{c0}$ is the critical temperature of clean sample). The scattering rate in principle can be extracted from the mobility, but, as discussed above, the multiband nature of these compounds does not allow a reliable quantitative evaluation of the carrier mobility. As an example for the *x* =0.36 sample with $T_c \approx 16$ K we find $g_{MR} \approx 0.5$ from $\mu_{MR}$ and $g_H \approx 30$ from $\mu_H$. An alternative evaluation can be extracted by considering the upper critical field. For the *x*=0.36 sample $\mu_0 |dH_{c2}/dT_c|/T_c \approx 0.35$ T/K is nearly twice the clean limit value (see the inset of Fig. 14 ) which means $\xi_0/l \approx 0.75$. The parameter *g* that can be written as $g = \hbar\Gamma/2\pi k_B T_{c0} \approx (T_c/T_{c0})\xi_0/l$ comes out to be $g_{Hc2} \approx 0.25$. Thus, the *g* values obtained with different criteria differ by two orders of magnitude. These differences lie on the crude evaluations that one necessary does in a multiband system and hinder any reliable comparison with theoretical models. However, if we reject $g_H$, extracted by $\mu_H$, that, as discussed above, is the most affected by the compensated nature of these compounds, it is possible to extrapolate the critical values $g_c$ at which $T_c$ is zero. We obtain $g_c \approx 0.7$ and $g_c \approx 0.3$ as extracted by MR and $H_{c2}$ respectively. These values are rather small in comparison with those estimated in irradiated Nd-1111[28] and La-1111[27], Co doped Nd-1111[10] and Co doped La-1111[10]. Indeed previous reports[10,27] extract the scattering rate from $\mu_H$ that strongly underestimates the actual mobility, while in Ref. 28 the carrier density is evaluated from the penetration depth. In the case of $s^\pm$ coupling rather small $g_c$ values are predicted,[3,10] that were ruled out by previous reports but cannot be excluded by our results. However, giving a reliable evaluation of the $g_c$ is out of the aim of this work. We rather focus on the trouble of extracting reliable transport parameters in condition of compensated compounds. This should be a warning for avoiding hasty conclusions on the nature of coupling without the chance to separate the contribution of different bands.

*iii) Role of Fe magnetism*

In order to explain the suppression of superconductivity in the Ru-rich phase (*x* >0.5) the role of Fe magnetic order in the parent compound should be considered. Ab-initio calculation shows that Ru atoms do not have any tendency to sustain a magnetic moment. These predictions agree very well with results obtained in Pr(Fe$_{1-x}$Ru$_x$)AsO samples (*x*≤0.75) in which the SDW ordering is observed up to *x*=0.67 and it disappears for *x* =0.75[14]. The disappearing of magnetic ordering in the parent compound coincides with the vanishing of superconductivity in the doped compound for *x*=0.75. If superconductivity is supposed to be related with spin fluctuations which survives after the suppression of SDW ordering, therefore magnetic order in the parent compounds becomes a fundamental prerequisite for the occurrence of superconductivity.

# 6. Conclusion

In this paper we present systematic investigations on disorder effects induced by isoelectronic substitutions in the FeAs layer of Sm-1111 family compounds on structural, transport, magnetic and superconducting properties. To this aim, a full series of $Sm(Fe_{1-x}Ru_x)As(O_{0.85}F_{0.15})$ samples with different Ru content was synthesized and first-principles calculations in the DFT framework were developed.

Detailed structure refinements clearly put in evidence that the cell parameter $a$ increases almost linearly with $x$, in good agreement with DFT calculation findings, whereas $c$ exhibits an almost constant value and decreases beyond $x \sim 0.1$. We find a good agreement between DFT calculations and experiments for all structural parameters with the exception of $c$ and Fe-As interplanar distance at low Ru content ($x \leq 0.25$), where a subtle interplay between magnetism and structure is present.

Transport properties in normal and superconducting state exhibit a rather complex behavior with increasing Ru substitution, which can be well rationalized thanks to the predictions of ab-initio calculations. Concerning the magnetic structure, DFT results show that Ru atoms do not sustain any magnetic moment and Ru substitution frustrates progressively Fe moments: a threshold for the stability of magnetism seems to be $x \sim 0.5$ and magnetic order is completely destroyed for $x > 0.75$. On the contrary, the electronic structure is only slightly affected by Ru substitution around the Fermi level: the most important effect is the broadening of Ru derived bands, which is a signature of a larger hybridization with As states. Small global charge doping is predicted, even in the presence of very large Ru contents. In all cases, $e$ and $h$ bands are found nearly compensated as inferred from Hall coefficient calculations.

These inputs allow to discuss the experimental results by considering separately Fe-rich ($x < 0.5$) and Ru-rich ($x > 0.5$) phases. In the former, transport properties are strongly affected by the disorder induced by Ru ions, which act as non magnetic impurities, whereas in the latter a metallic behavior is recovered due to the increasing bandwidth of Ru-rich compositions. Superconducting properties can be understood within this framework: in the Fe-rich phase, Ru substituted samples move progressively from clean to dirty limit and $T_C$ is suppressed by pair-breaking impurity scattering. The compensated nature of these compounds avoid a reliable evaluations of the critical scattering rate at which superconductivity is completely suppressed. Nevertheless our estimations do not rule out that $T_c$ might be suppressed by rather weak not magnetic scattering as predicted for unconventional s± pairing.

Notwithstanding the effects of disorder are partially healed in the Ru-rich phase, $T_C$ vanishes for $x \geq 0.75$ that is a rather large value in comparison with other substitutions in the FeAs layer (Mn and Co), where band filling completely suppresses superconductivity for 10% substitution. DFT calculations show that substitution of Fe with Ru induces no appreciable band filling effects and therefore it preserves the existence of compensated $h$ and $e$ sheets, even in the Ru-rich phase. On the other hand for $x \geq 0.75$ magnetic order in the parent compound is thought to be suppressed. Both these observations suggest that two main ingredients should be simultaneously present to establish and/or preserve superconductivity in pnictides: *i*) compensated $h$ and $e$ bands and *ii*) magnetic correlations survived after SDW order suppression. The occurrence of both these conditions suggests that an unconventional pairing coupling interaction could be the origin of superconductivity in this new class of superconductors.

**Acknowledgments**

This work is partially supported by Compagnia di S. Paolo and by the European Commission from the 7th framework program "Transnational Access," under Contract No. 228043-EuroMagNETII – Integrated Activities and by the Italian Foreign Affairs Ministry (MAE) - General Direction for the Cultural Promotion. The authors are grateful to E. Mossang for his help at GHMFL.